\documentclass[useAMS,usenatbib,usegraphicx]{mn2e}

\usepackage{aas_macros}

\makeatletter
\DeclareRobustCommand{\text}{%
  \ifmmode\expandafter\text@\else\expandafter\mbox\fi}
\let\nfss@text\text
\def\text@#1{{\mathchoice
  {\textdef@\displaystyle\f@size{#1}}%
  {\textdef@\textstyle\f@size{#1}}%
  {\textdef@\textstyle\sf@size{#1}}%
  {\textdef@\textstyle \ssf@size{#1}}%
  \check@mathfonts
  }%
}
\def\textdef@#1#2#3{\hbox{{%
                    \everymath{#1}%
                    \let\f@size#2\selectfont
                    #3}}}
\makeatother

\title[Sodium absorption in WASP-17b]{Detection of sodium absorption in WASP-17b with Magellan\thanks{This paper includes data gathered with the 6.5 meter Magellan Telescopes located at Las Campanas Observatory, Chile.}}
\author[G. Zhou and D. D. R. Bayliss]{G. Zhou$^{1}$\thanks{E-mail: george@mso.anu.edu.au} and D. D. R. Bayliss$^{1}$\\
$^{1}$Research School of Astronomy and Astrophysics, Australian National University, Cotter Rd, Weston Creek, ACT 2611, Australia}

\begin{document}

\date{Accepted 2012 July 28.  Received 2012 July 13; in original form 2012 June 1}

\pagerange{\pageref{firstpage}--\pageref{lastpage}} \pubyear{2012}

\maketitle

\label{firstpage}

\begin{abstract}
We present the detection of sodium absorption in the atmosphere of the extrasolar planet WASP-17b, an inflated `hot-Jupiter' in a tight orbit around an F6 dwarf. In-transit observations of WASP-17 made with the MIKE spectrograph on the 6.5-m Magellan Telescope were analysed for excess planetary atmospheric absorption in the sodium I `D' doublet spectral region. Using the interstellar sodium absorption lines as reference, we detect an excess $0.58 \pm 0.13\, \text{per cent}$ transit signal, with $4.5 \sigma$ confidence, at 1.5 \AA\, bandwidth around the stellar sodium absorption feature. This result is consistent with the previous VLT detection of sodium in WASP-17b, confirming that the planet has a highly inflated atmosphere. 
\end{abstract}


\section{Introduction}
\label{sec:intro}

Numerous techniques have now been developed to study the atmospheric properties of transiting planets. For example, emission from the planet inferred by observations of the secondary eclipse can reveal their day-side temperatures \citep[e.g.][]{Charbonneau2005,Deming2005}. Additional infrared photometry of the phase variation can map the temperature distribution of `hot-Jupiter' planets \citep{Knutson2007}. As light from the host star passes through the upper atmosphere of a planet during the primary transit, it is possible to investigate the planetary atmospheric composition via transmission spectroscopy. \citet{Charbonneau2002} employed this technique to detect the atmosphere of an extrasolar planet for the first time. Their observations of HD 209458b with the Hubble Space Telescope (HST) revealed a 0.02 per cent transit signal in the NaI resonance lines, indicative of NaI absorption in the upper atmosphere of the planet.

The amount of absorption measured by transmission spectroscopy probes the wavelength dependence of the height of the planet's atmosphere. The optical depth is dependent on the temperature, surface gravity, and composition of the atmosphere. For `hot-Jupiters', the opacity is predicted to be strongest at the NaI and KI resonance lines in the optical, whilst molecular species such as water and methane dominate the near-infrared part of the spectrum \citep{Seager2000, Brown2001,Hubbard2001}. In recent years, the use of transmission spectroscopy have lead to a series of discoveries in the study of `hot-Jupiter' atmospheres, including the escaping atomic hydrogen envelope around HD 209458 \citep{Vidal2003}, and the presence of haze in HD 189733b \citep{Pont2008}.

Despite the recent successes, the expected planetary absorption signal is very small, and the technique of transmission spectroscopy remains difficult. So far, NaI detection via ground based high-resolution spectroscopy has only been achieved on three transiting planets, HD 209458b \citep{Snellen2008,Langland2009,Jensen2011}, HD 189733b \citep{Redfield2008}, and WASP-17b \citep{Wood2011}, along with numerous unsuccessful attempts \citep[e.g.][]{Narita2005,Bozorgnia2006}. Recent detections around the KI resonance line by narrowband photometry has also been achieved with XO-2b \citep{Sing2011} and HD 80606b \citep{Colon2012}. High-resolution spectroscopic detections often require 8-10 m class telescopes, with data taken in optimal conditions, yielding signal-to-noise ratios $\text{S/N} \gg 100 \, \text{pixel}^{-1}$.  Even so, the transit signal is heavily contaminated by much larger systematic trends, such as echelle blaze function variations \citep{Winn2004}, CCD non-linearity \citep{Snellen2008}, and telluric contamination \citep[e.g.][]{Albrecht2009,Langland2009,Jensen2011}. Here, we present a novel approach of using the nearby interstellar absorption lines to correct for the majority of these systematic effects. 

WASP-17b \citep{Anderson2010} is a $0.49\, M_\text{Jup}$, $1.99\, R_\text{Jup}$ planet in a retrograde, 3.7 day orbit \citep{Bayliss2010,Triaud2010}. The equilibrium temperature of the planet, measured by \emph{Spitzer} photometry of the secondary eclipse, is $1771\pm35 \, \text{K}$, assuming zero albedo and full redistribution \citep{Anderson2011}. Recent observations by \citet{Wood2011} with GIRAFFE on the VLT detected a 1.46 per cent transit in the NaI `D' lines at $0.75 \, \text{\AA}$ bandwidth, a level ten times larger than detections made around any other planet. 

\citet{Bayliss2010} obtained time-series spectra of WASP-17 during the primary transit event to measure the spin-orbit alignment of the system. In this study, we present an analysis of this data to monitor for NaI absorption in the planetary atmosphere.

\section{Observations and Data Analysis}
\label{sec:observ-data-analys}

Observations of WASP-17 were obtained with the Magellan Inamori Kyocera Echelle (MIKE) spectrograph on the 6.5-m Magellan II (Clay) telescope, located at Las Campanas Observatory, on the night of 2010 May 11. The spectra were obtained at a resolution of $\lambda / \Delta \lambda \approx 48000$ in the wavelength range 5000-9500 \AA,\ over echelle orders 37-69. Each exposure was $600\,\text{s}$ in length, giving an average S/N of $82\,\text{pixel}^{-1}$  over the NaI `D' doublet region, in order 58. Th-Ar arc lamp exposures were taken bracketing each exposure. Of the 37 object exposures, 21 were taken out-of-transit, 11 in-transit, 2 during ingress, and 3 during egress. Two observations were rejected for having $\text{S/N} < 60$, one was taken during egress, the other whilst out-of-transit near the end of the night, at high airmass. Observations were paused for 30 minutes, whilst the planet was in-transit, due to the zenith limit of the telescope. The details of the observations and data reduction are given in \citet{Bayliss2010}.

\subsection{Spectral analysis}
\label{sec:spectral-analysis}

\subsubsection{Correction of systematic trends}
\label{sec:corr-syst-trends}

\begin{figure*}
\begin{minipage}{17cm}
  \centering
  \includegraphics[width=16cm]{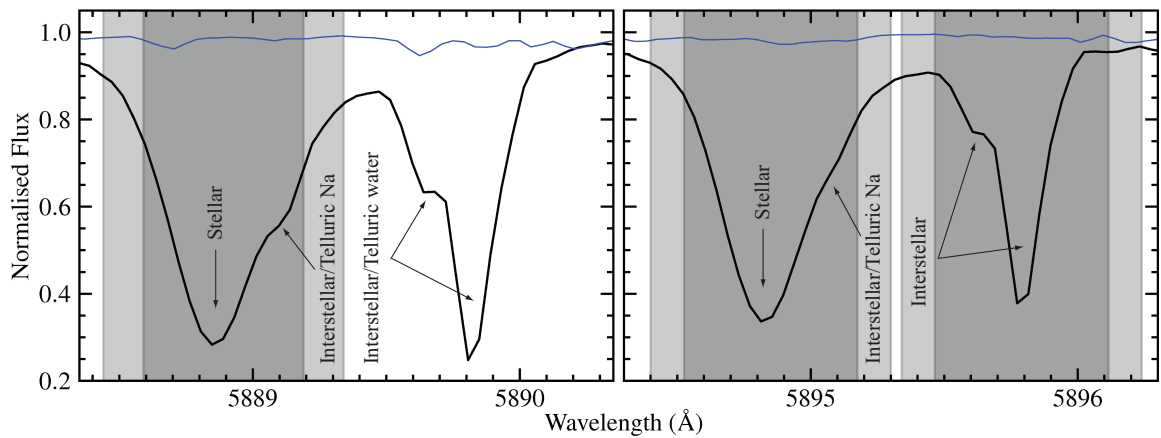}
  \caption{Plotted are the NaI `D' doublet regions of the template spectrum, with D$_1$ on the left panel, D$_2$ on the right panel, as observed with Magellan. The broad stellar absorption features are resolved from the narrower interstellar components. Each line is also blended with additional interstellar and telluric NaI components \citep{Wood2011}. The minimum and maximum half-bandwidths of 0.60 and 0.90 \AA\, are filled in grey. The telluric absorption spectrum, as observed with a rapidly rotating B star, is over-plotted in blue.}
  \label{fig:NaD_spectrum}
\end{minipage}
\end{figure*}

A wavelength shift of $\sim1\,\text{km}\,\text{s}^{-1}$ was registered over the night, due to a combination of instrument drift, airmass change, system orbital dynamics, and the Rossiter-McLaughlin (RM) effect \citep{Rossiter1924,McLaughlin1924}. The average shift per exposure was calculated by weighted averaging over all orders not contaminated by telluric absorption, and corrected for with the IRAF\footnote{IRAF is distributed by the National Optical Astronomy Observatories, which are operated by the Association of Universities for Research in Astronomy, Inc., under cooperative agreement with the National Science Foundation} package \emph{RV}\@. We found time and wavelength dependent variations in the blaze function between exposures, and corrected for this using the method described in \citet{Winn2004}. We fitted the residual difference between an observed spectrum and an averaged template with an order 5 smoothed B-spline. In order to avoid removing the actual transit signal in the analysis, we corrected the spectrum of order $n$ with the averaged residual fits of orders $n-1$ and $n+1$. Finally, a template spectrum was created by a $2\sigma$ clipped averaging of all blaze corrected object spectra. The NaI `D' doublet region of the template spectrum is shown in Fig.~\ref{fig:NaD_spectrum}.

\subsubsection{Subtraction of telluric features}
\label{sec:subtr-tell-feat}

Telluric absorption lines originating from water and Na are dispersed unevenly within the stellar and interstellar NaI features (Fig~\ref{fig:NaD_spectrum}). Nightly variations in the strength of these telluric lines can overwhelm the planetary NaI absorption signal, and must be removed in the spectral analysis process. Some transmission spectroscopy observations \citep[e.g.][]{Winn2004,Narita2005,Redfield2008} removed their telluric contamination by subtracting a smooth B star spectrum; others \citep[e.g.][]{Snellen2008,Wood2011} performed the same subtraction using synthetic telluric spectra. We obtained a telluric template spectrum by averaging two observations of the B star HD 129116. For each WASP-17 observation, the telluric spectrum, scaled and shifted by fitting for the telluric water dominated regions around the NaI `D' doublet, was divided from the object spectrum. We note that the telluric scaling deviates from the airmass function at the start of the night, indicative of variations in the atmospheric water content. The telluric scaling correlates well with airmass for the second half of the night. We see no significant telluric Na lines in the telluric spectrum, and assume that any telluric Na variations scale with water. Similarly, \citet{Snellen2008} pointed out that although telluric Na varies seasonally, it correlates well with water and oxygen over a single night.

\subsubsection{Flux extraction}
\label{sec:flux-extraction}

Following Equation 1 of \citet{Snellen2008}, we integrated the flux within a half-bandwidth of the line centroid, corrected against two adjacent reference continuum regions. We found that the blaze and reference continuum region corrections could not remove all systematic effects, but variations in the interstellar NaI lines correlate strongly with that of the stellar NaI lines. The strong interstellar NaI absorption doublet can be found $1\, \text{\AA}$ $(\sim 50 \, \text{km} \, \text{s}^{-1})$ red-ward of their stellar components (Fig.~\ref{fig:NaD_spectrum}). We can assume that the strength of the interstellar absorption features are invariant with time, as we do not expect the interstellar medium to be changing on the time scale of hours. The proximity of the interstellar to the stellar features makes them excellent references to correct for any additional, higher order systematic variations. We hence extract the flux of both the stellar and interstellar absorption lines in our analysis.

The NaI flux was calculated as the ratio between the sum of the contributing fluxes from each component of the doublet and that of the same regions in the template spectrum. The line centroids of the stellar NaI `D' doublet are located at 5888.89 (D$_1$) and 5894.85 \AA\, (D$_2$). The adjacent interstellar components are located at 5889.82 and 5895.80 \AA. 

We find that the interstellar D$_1$ line is severely contaminated by telluric water lines (Fig~\ref{fig:NaD_spectrum}). The line shows large time dependent variations, most likely due to the incomplete subtraction of telluric lines blended within the stellar absorption feature. The relative level of telluric contamination can be quantified by taking the sum equivalent width of the telluric lines within the stellar and interstellar regions; for the stellar D$_1$ and D$_2$ lines the sum telluric equivalent widths are 0.0046 \AA\, and 0.0043 \AA, for the interstellar D$_1$ and D$_2$ lines the sum telluric equivalent widths are 0.0092 \AA\, and 0.0008 \AA. The interstellar D$_1$ line contains more than twice the amount of telluric contamination as the stellar D$_1$, D$_2$ and the interstellar D$_2$ lines. We therefore do not use the interstellar D$_1$ line in our subsequent analyses. 

For each exposure, we performed a Gaussian centroid fit to each line to remove any uncorrected wavelength shifts, then summed the flux around the fitted centroid position. The reference continuum regions to the left and right of the line centroid, for the stellar NaI D$_1$ line, are centred at 5887 \AA\, and 5893 \AA\, respectively, and for the stellar and interstellar NaI D$_2$ lines at 5893 and 5897 \AA\, respectively. These reference regions are not located directly adjacent of the stellar lines to avoid overlapping with the interstellar components.

The use of the nearby interstellar NaI absorption lines as reference is facilitated by the large velocity difference between WASP-17 and the sight line interstellar medium. The heliocentric mean velocity of WASP-17 is $-49.5\, \text{km}\,\text{s}^{-1}$ \citep{Anderson2010}, whilst we measure the sight line interstellar medium velocity to be $-0.5\, \text{km}\,\text{s}^{-1}$, consistent with the velocity of the local interstellar cloud \citep[e.g.][]{Redfield2008b}. The separation between the stellar and interstellar NaI absorption features is $\sim 1\, \text{\AA}$ in wavelength space. To investigate the minimum stellar--interstellar feature separation required for this technique, we simulate a stellar, interstellar absorption pair with two equal height Gaussians of FWHM 0.4 and 0.2 \AA. Various separations between the two Gaussians are tested, and the flux is extracted about the broader stellar Gaussian at 0.6 \AA\, bandwidth. We find that at separations less than 0.54 \AA, the influence of the interstellar line on the extracted stellar line flux becomes greater than our measurement error of 0.15\%, corresponding to a minimum velocity separation requirement of $27.5\, \text{km}\,\text{s}^{-1}$ between the target and the sight line interstellar medium.

For a single exposure, the NaI flux, $F_\text{rel}$, relative to the ratio between the stellar fluxes $F_\text{D1,stellar}$, $F_\text{D2,stellar}$ and the interstellar flux $F_\text{D2,interstellar}$, is:
\begin{equation}
  \label{eq:Naflux}
  F_\text{rel} = \frac{w_1 \cdot F_\text{D1,stellar} + w_2 \cdot F_\text{D2,stellar}}{F_\text{D2,interstellar}} \, .
\end{equation}
The weights $w_1$ and $w_2$ are inversely proportional to the scatter of the lightcurves calculated for each stellar line individually, according to
\begin{equation}
  \label{eq:Naflux_D1}
  F_\text{D1,rel} = \frac{F_\text{D1,stellar}}{F_\text{D2,interstellar}}
\end{equation}
and
\begin{equation}
  \label{eq:Naflux_D2}
  F_\text{D2,rel} = \frac{F_\text{D2,stellar}}{F_\text{D2,interstellar}} \, .
\end{equation}

The flux of each individual component is calculated by their relative counts to the same regions in the template spectrum, corrected by the counts in the reference continuum regions. For example, the flux in the stellar $D_1$ line is derived from the counts in the object spectrum about the absorption line and reference region, $O_\text{D1}$ and $O_\text{ref}$, and the counts in the same regions in the template spectrum, $T_\text{D1}$ and $T_\text{ref}$, by,
\begin{equation}
  \label{eq:D1flux}
  F_\text{D1,stellar} = \frac{O_\text{D1} / O_\text{ref}}{T_\text{D1} / T_\text{ref}} \, .
\end{equation}
The error in the relative flux is derived from the average S/N within the bandwidth, divided by the square root of the number of spectral points included in the bandwidth.

\subsubsection{Modelling of in-transit line profile variations}
\label{sec:modell-trans-line}

The strength of the absorption signal is strongly dependent on the bandwidth within which the flux is summed; the relationship between the bandwidth and transit depth provides information on the properties of the planetary atmosphere. One limitation with our methodology of using the interstellar NaI lines as calibrators is that we must restrict the size of the bandwidth used. To avoid overlapping between the stellar and interstellar NaI components, the maximum half-bandwidth we employ is 0.90 \AA. The minimum bandwidth of 0.60 \AA\, is determined by the line profile variations in-transit that results from the distortions induced by the RM effect.

The observed line profile is a convolution between the intrinsic stellar absorption profile, a time varying rotational kernel that takes into account the occulted stellar disc, and the instrument response kernel. To model the distortions, the intrinsic NaI stellar absorption profile was approximated using a high-resolution solar spectrum \citep{Delbouille1990}. Solar rotational broadening is four times less than that of WASP-17, whilst the instrumental broadening in the solar template is 100 times less than our observations. \citet{Hirano2010} provided a Gaussian approximation to the analytical solution of the rotation kernel. Using theoretical $V$ band limb darkening coefficients from \citet{Claret2011}, with \citet{Anderson2011} stellar parameters for WASP-17, we calculated a Gaussian rotational kernel with a standard deviation of $0.12 \, \text{\AA}$. The kernel is modified at each time step to reflect the position of the planet over the stellar disc during transit. We modelled the MIKE instrument kernel with a Gaussian of 0.052 \AA\, standard deviation, matching a spectral resolution of $\lambda / \Delta \lambda = 48000$ at 5890 \AA. The solar template is convolved at each time step, and the flux around the absorption line is summed and normalised against the out-of-transit flux.

\begin{figure}
  \centering
  \includegraphics[width=8cm]{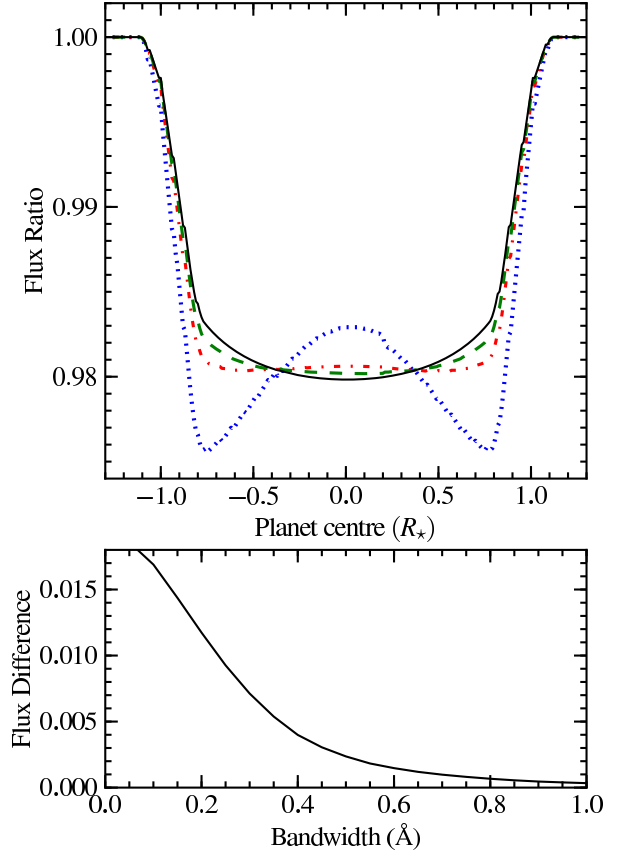}
  \caption{The top panel plots the model transit lightcurves from fluxes extracted around the NaI D$_1$ line at bandwidths of 2.0 \AA\, (black solid), 0.7 \AA\, (green dashed), 0.5 \AA\, (red dot-dashed), and 0.3 \AA\, (blue dotted). The bottom panel shows the absolute flux difference for the narrow bandwidths lightcurves, relative to a 2.0 \AA\, bandwidth lightcurve.}
  \label{fig:bandwidth_model_lc}
\end{figure}

The upper panel of Fig.~\ref{fig:bandwidth_model_lc} shows the resulting transit lightcurve derived from the flux around NaI D$_1$ at various bandwidths. The distortions to the transit lightcurve, calculated as the absolute difference to the $2.0\, \text{\AA}$ bandwidth lightcurve, become larger than our measurement error of 0.15 per cent at bandwidths narrower than 0.60 \AA\, (Fig.~\ref{fig:bandwidth_model_lc}, lower panel). 

\begin{figure}
  \centering
  \includegraphics[width=8cm]{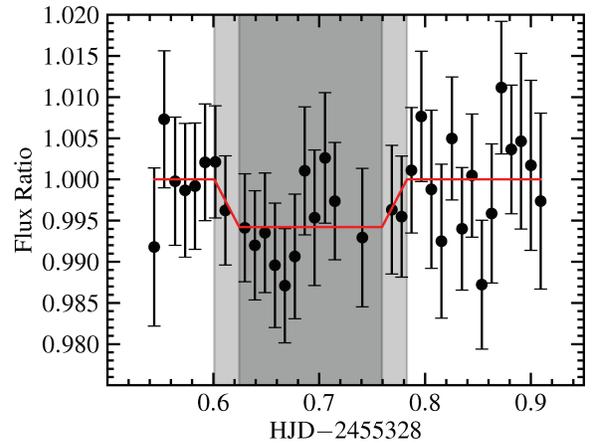}
  \caption{An example lightcurve, extracted around the stellar NaI D$_1$ and D$_2$ lines at 0.75 \AA\, half-bandwidth, is plotted. The red line plots the $0.58 \pm 0.13 \, \text{per cent}$ transit depth fit. The darker shade marks the in-transit proportion of the lightcurve, whilst the lighter shade marks ingress and egress.}
  \label{fig:lightcurve}
\end{figure}

We therefore extracted the flux around the NaI `D' doublet at seven half-bandwidths, from 0.60 to 0.90 \AA, evenly spaced at 0.05 \AA\, intervals. As an example, the 0.75 \AA\, half-bandwidth lightcurve is plotted in Fig~\ref{fig:lightcurve}, and the fluxes are set out in Table~\ref{tab:075_flux_table}.

\begin{table}
  \caption{Fluxes for the $0.75 \, \text{\AA}$ half-bandwidth lightcurve}
  \centering
  \begin{tabular}{cccc}
    \hline
    $\text{HJD}-2455328$ & $F_\text{rel}$ & $\Delta F_\text{rel}$ & Comment\\
    \hline
0.544 & 0.9918 & 0.0096 & Out-of-transit \\
0.554 & 1.0073 & 0.0083 & Out-of-transit \\
0.564 & 0.9998 & 0.0078 & Out-of-transit \\
0.573 & 0.9987 & 0.0081 & Out-of-transit \\
0.583 & 0.9992 & 0.0077 & Out-of-transit \\
0.592 & 1.0021 & 0.0071 & Out-of-transit \\
0.602 & 1.0021 & 0.0068 & Ingress \\
0.611 & 0.9962 & 0.0066 & Ingress \\
0.630 & 0.9941 & 0.0066 & In-transit \\
0.639 & 0.9920 & 0.0066 & In-transit \\
0.649 & 0.9935 & 0.0073 & In-transit \\
0.658 & 0.9896 & 0.0076 & In-transit \\
0.668 & 0.9871 & 0.0069 & In-transit \\
0.677 & 0.9906 & 0.0076 & In-transit \\
0.686 & 1.0011 & 0.0078 & In-transit \\
0.696 & 0.9954 & 0.0082 & In-transit \\
0.706 & 1.0026 & 0.0079 & In-transit \\
0.715 & 0.9974 & 0.0071 & In-transit \\
0.741 & 0.9929 & 0.0084 & In-transit \\
0.769 & 0.9963 & 0.0078 & Egress \\
0.778 & 0.9955 & 0.0074 & Egress \\
0.787 & 1.0011 & 0.0076 & Out-of-transit \\
0.796 & 1.0077 & 0.0079 & Out-of-transit \\
0.806 & 0.9988 & 0.0096 & Out-of-transit \\
0.815 & 0.9925 & 0.0093 & Out-of-transit \\
0.825 & 1.0050 & 0.0075 & Out-of-transit \\
0.835 & 0.9940 & 0.0074 & Out-of-transit \\
0.844 & 1.0005 & 0.0075 & Out-of-transit \\
0.854 & 0.9872 & 0.0078 & Out-of-transit \\
0.863 & 0.9959 & 0.0084 & Out-of-transit \\
0.872 & 1.0112 & 0.0081 & Out-of-transit \\
0.881 & 1.0036 & 0.0078 & Out-of-transit \\
0.891 & 1.0046 & 0.0107 & Out-of-transit \\
0.900 & 1.0017 & 0.0103 & Out-of-transit \\
0.909 & 0.9974 & 0.0107 & Out-of-transit \\
    \hline
  \end{tabular}
  \label{tab:075_flux_table}
\end{table}

\subsection{Lightcurve analysis}
\label{sec:lightcurve-analysis}

The transit depth, $D$, is the ratio of the average out-of-transit flux to the average in-transit flux, $F_\text{OOT}$ and $F_\text{INT}$ respectively \citep{Brown2001},
\begin{equation}
  \label{eq:Brown2011_depth}
  D = 1 - \frac{F_\text{INT}}{F_\text{OOT}} \, .
\end{equation}
Exposures with mid-times set during ingress or egress are excluded. Overall, there were 20 out-of-transit points and 11 in-transit points included in the calculations. The error in the transit depth, $\Delta D$, is estimated from the standard deviation of the points in and out-of-transit, $\sigma_\text{INT}$ and $\sigma_\text{OOT}$ respectively, 
\begin{equation}
  \label{eq:errors}
  \left( \frac{\Delta D}{1-D} \right)^2 = \left( \frac{\sigma_\text{INT} / \sqrt{N_\text{INT}}}{F_\text{INT}} \right)^2 + \left( \frac{\sigma_\text{OOT} / \sqrt{N_\text{OOT}}}{F_\text{OOT}} \right)^2 \, ,
\end{equation}
where $N_\text{INT}$ and $N_\text{OOT}$ are the number of points averaged in and out-of-transit. The same analysis was performed at every bandwidth, the results of which are presented in Table~\ref{tab:bandwidth_transitdepth}. At 1.5 \AA\, bandwidth, we detect a transit depth of $0.58 \pm 0.13 \, \text{per cent}$ with $4.5\sigma$ confidence. We also performed the same lightcurve analysis on the fluxes extracted around the stellar NaI D$_1$ and D$_2$ lines individually, according to Eq~\ref{eq:Naflux_D1} and \ref{eq:Naflux_D2}. The transit depths at 1.5 \AA\, bandwidth for the D$_1$ and D$_2$ stellar lines are $0.53 \pm 0.16$ and $0.66 \pm 0.10 \, \text{per cent}$, respectively. The transit depths derived from each individual NaI line is consistent with each other and with that derived from the weighted average lightcurve.

\begin{table}
  \caption{The transit depths at various bandwidths.}
  \centering
  \begin{tabular}{cc}
    \hline
    Bandwidth (\AA) & Transit depth (per cent)\\
    \hline
    1.2 & $0.60 \pm 0.14$\\
    1.3 & $0.60 \pm 0.13$\\
    1.4 & $0.59 \pm 0.13$\\
    1.5 & $0.58 \pm 0.13$\\
    1.6 & $0.56 \pm 0.13$\\
    1.7 & $0.50 \pm 0.12$\\
    1.8 & $0.42 \pm 0.11$\\
    \hline
  \end{tabular}
  \label{tab:bandwidth_transitdepth}
\end{table}

\section{Discussion}
\label{sec:discussion}

Our Magellan transmission spectroscopic measurements detected a 0.58 per cent transit, at $4.5 \sigma$ confidence, in the NaI `D' doublet region of WASP-17, which is consistent with the results of \citet{Wood2011}. This study confirms that WASP-17b shows the strongest NaI absorption signal measured to date. Given the low surface gravity and high equilibrium temperature of the planet, WASP-17b has the largest atmospheric scale height of all known transiting planets, facilitating the large transmission spectroscopic signal.

Interestingly, an extension of our analysis to narrower bandwidths could not replicate the steep increase in absorption strength as observed by \citet{Wood2011}. It is unclear how the RM effect distortions to the absorption line profile were dealt with in this previous study, given that fluxes extracted at 0.375 \AA\, half-bandwidth would have been severely affected. In addition, our results hint at a steep decrease in absorption strength for our broader bandwidths. This feature is potentially similar to the steep decrease observed by \citet{Wood2011} at 3.0 \AA\, bandwidth, which was attributed primarily to the presence of clouds in the upper planetary atmosphere.

\begin{figure}
  \centering
  \includegraphics[width=8cm]{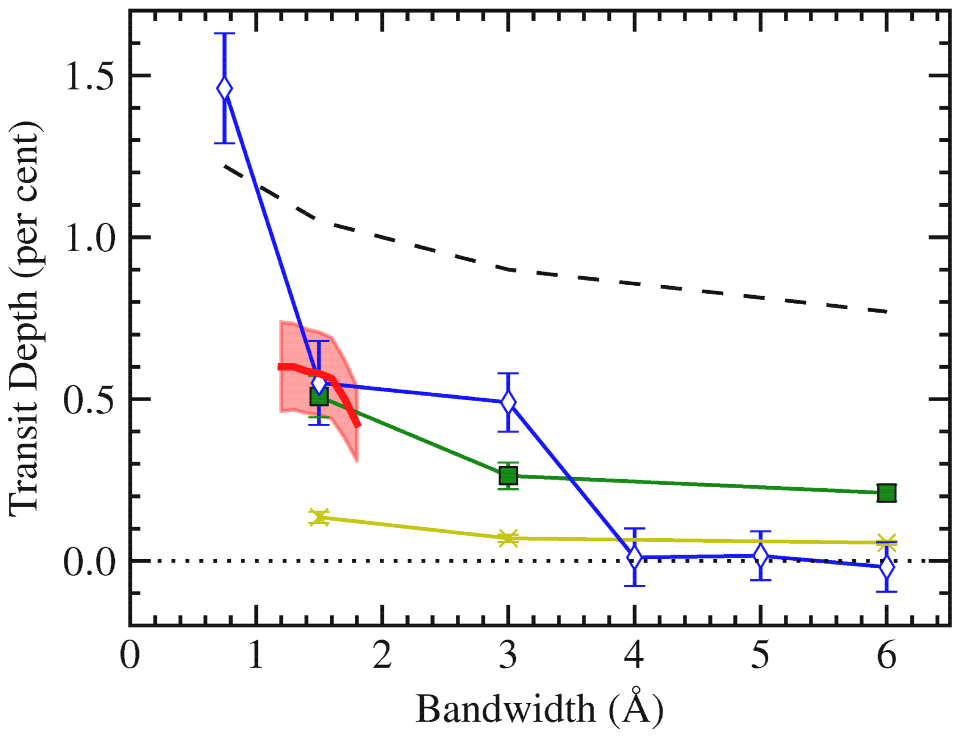}
  \caption{Plotted are the various observed and modelled NaI transit depths. The Magellan detection presented in this paper is shown by red shaded region. Blue diamonds plot the results of \citet{Wood2011}. Green squares plot HD 209458b measurements from \citet{Snellen2008} after scaling by the WASP-17b scale height. Yellow crosses plot the original \citet{Snellen2008} measurements of HD 209458b. The \citet{Brown2001} model prediction, as calculated and scaled by \citet{Wood2011}, is plotted as the dashed line.}
  \label{fig:band_depth}
\end{figure}

We can make a comparison between observations of HD 209458b and that of WASP-17b by scaling the former according to the differences in temperature and gravity between the two planets. The transit depth is governed by the atmospheric scale height of the planet and the radius ratio between the planet $(R_P)$ and the host star $(R_\star)$ \citep{Brown2001}:
\begin{equation}
  \label{eq:transit_depth}
  D = \frac{2\pi R_P (kT/\mu g)}{\pi R_\star^2} \, ,
\end{equation}
where $T$ is the temperature, $\mu$ is the mean molecular weight of the atmosphere, and $g$ is the surface gravity of the planet. Such scaling assumes that the atmospheric structure is consistent between the planets, and the change in contribution from secondary effects are negligible, accounting only for the difference in temperature and bulk density between the two planets. Taking a scale height for HD 209458b of 500 km and $\mu$ of H$_2$, we multiply the \citet{Snellen2008} measurements by a factor of 2.4, the resulting estimates are plotted in Fig.~\ref{fig:band_depth}. Our WASP-17b measurements match the scaled HD 209458b atmosphere remarkably well.

Similarly, the transit depths derived and scaled from models of \citet{Brown2001} by \citet{Wood2011} are also plotted. The model prediction lies well above our measured signal. All transmission spectroscopy of transiting planets to date have consistently shown that the NaI signal is weaker than that of a cloudless model in local thermodynamic equilibrium. The close-in orbits of these `hot-Jupiters' mean that their atmospheres are highly irradiated, causing depletion of Na to form Na$^+$. Photoionisation occurs in the upper levels of the atmosphere, whilst the line wings are formed in the denser lower atmosphere. The result of neutral Na depletion is therefore a reduction in the depth of the line cores \citep[e.g.][]{Barman2002,Fortney2003}. The presence of silicate and iron condensates in the upper atmosphere \citep{Sudarsky2000,Sudarsky2003,Fortney2003}, combined with the slanted geometry inherent to transmission spectroscopy \citep{Fortney2005}, also act to decrease the absorption signal. 

We note that this study presents a ground based spectroscopic NaI detection with the smallest telescope to date, in a S/N regime significantly lower than previous observations. This shows that the use of interstellar absorption features as reference is a robust technique that removes the majority of systematic effects. \citet{Langland2009} noted that better telluric subtractions could be achieved by taking interleaved smooth spectrum star observations, reducing the reliance on scaling a single telluric template to fit the observed spectrum. WASP-17 is by far the faintest target pursued by the transmission spectroscopy technique, the positive detection opens up the possibility to explore other similarly inflated targets.

\section*{Acknowledgments}

Australian access to the Magellan Telescopes was supported through the National Collaborative Research Infrastructure Strategy of the Australian Federal Government. D.D.R.B acknowledges financial support from the Access to Major Research Facilities Programme, which is a component of the International Science Linkages Programme established under the Australian Government innovation statement, Backing Australias Ability. We thank J.N. Winn, R.A. Mardling, and P.D. Sackett for contributions to the original data. We also thank D. Yong and C. Fishlock for useful discussions and for supplying the MIKE telluric spectrum.

\bibliographystyle{mn2e}
\bibliography{mybibfile}

\label{lastpage}

\end{document}